\documentclass{emulateapj}

\usepackage{amsmath}
\usepackage{selectp}
\usepackage{mathtools,amssymb}

\usepackage{savesym}
\usepackage{multirow}
\savesymbol{longtable*}
\usepackage{subfig}
\restoresymbol{SF}{longtable*}

\begin{document}
\slugcomment{Close Interstellar Comets}

\title{Realistic Detectability of Close Interstellar Comets}
\author{Nathaniel V. Cook\altaffilmark{1}, Darin Ragozzine\altaffilmark{2,3,4}, Mikael Granvik\altaffilmark{5,6}, Denise C. Stephens\altaffilmark{1}}
\email{nathanielvcook@gmail.com}
\altaffiltext{1}{Brigham Young University, BYU Department of Physics and Astronomy N283 ESC, Provo, UT 84602, USA}
\altaffiltext{2}{Harvard-Smithsonian Center for Astrophysics, 60 Garden Street, Cambridge, MA 02138, USA}
\altaffiltext{3}{University of Florida, 211 Bryant Space Science Center, Gainesville, FL 32611, USA}
\altaffiltext{4}{Florida Institute of Technology, 150 West University Boulevard, Melbourne, FL 32901, USA}
\altaffiltext{5}{Department of Physics, P.O.\ Box 64, 00014 University of Helsinki, Finland}
\altaffiltext{6}{Finnish Geospatial Research Institute, P.O.\ Box 15, 02430 Masala, Finland}

\begin{abstract}
During the planet formation process, billions of comets are created and ejected into interstellar space. The detection and characterization of such interstellar comets (also known as extra-solar planetesimals or extra-solar comets) would give us \emph{in situ} information about the efficiency and properties of planet formation throughout the galaxy. However, no interstellar comets have ever been detected, despite the fact that their hyperbolic orbits would make them readily identifiable as unrelated to the solar system. Moro-Mart{\'{\i}}n et al. 2009 have made a detailed and reasonable estimate of the properties of the interstellar comet population. We extend their estimates of detectability with a numerical model that allows us to consider ``close'' interstellar comets, e.g., those that come within the orbit of Jupiter. We include several constraints on a ``detectable'' object that allow for realistic estimates of the frequency of detections expected from the Large Synoptic Survey Telescope (LSST) and other surveys. The influence of several of the assumed model parameters on the frequency of detections is explored in detail. Based on the expectation from Moro-Mart{\'{\i}}n et al. 2009, we expect that LSST will detect 0.001-10 interstellar comets during its nominal 10-year lifetime, with most of the uncertainty from the unknown number density of small (nuclei of $\sim$0.1-1 km) interstellar comets. Both asteroid and comet cases are considered, where the latter includes various empirical prescriptions of brightening. Using simulated LSST-like astrometric data, we study the problem of orbit determination for these bodies, finding that LSST could identify their orbits as hyperbolic and determine an ephemeris sufficiently accurate for follow-up in about 4--7 days. We give the hyperbolic orbital parameters of the most detectable interstellar comets. Taking the results into consideration, we give recommendations to future searches for interstellar comets. 
\end{abstract}

\keywords{circumstellar matter -- comets: general -- Kuiper Belt -- minor planets, asteroids -- planetary systems -- solar system: formation}

\maketitle

\section{Introduction}

The current understanding of planet formation suggests that very large numbers of minor bodies are ejected into interstellar space by planets during and after formation \citep[e.g.,][]{1972IAUS...45..329S,1987AJ.....94.1330D,2004come.book..153D}. In typical simulations of solar system formation, only a small fraction of the small bodies that have a close encounter with the giant planets are captured into the Oort cloud \citep[the current source of long-period comets,][]{Oort1950}, the rest are ejected into the interstellar medium. Despite changing understanding of the formation and properties of our Oort cloud \citep[e.g.,][]{Levison2010,2011Icar..215..491K}, extra-solar debris disks \cite[e.g.,][]{2012Natur.490...74D}, and planet formation \citep[e.g.,][]{2012arXiv1211.1673C}, there is general consensus that interstellar space must be populated with a non-trivial population of small bodies, including those corresponding in size to the asteroids and comets in the solar system. 

Minor planets that originate in other planetary systems but are currently unbound are usually called interstellar comets (ICs)\index{interstellar comet}.\footnote{Although there are no codified definitions, objects unbound from any star that are the natural minor body extension of the interstellar medium are usually called interstellar comets \citep[e.g.,][]{1975AJ.....80..525W,1976Icar...27..123S}, while minor bodies detected orbiting around other stars are a natural extension of extra-solar planets and are often referred to as extra-solar planetesimals\citep[e.g.,][]{2005AJ....130.1261J}. \citet{M09} is an exception to this typical nomenclature.} Although there is one candidate IC known ($\S$\ref{sec:candidate}), at present, their existence is essentially theoretical \citep{1975AJ.....80..525W,1976Icar...27..123S, McGlynn89, Francis05}. Technically, many long period comets have slightly hyperbolic orbits, but these clearly originate in the solar system and only appear unbound at present due to minor gravitational and non-gravitational perturbations, so are not considered ICs. Indeed, identification of an object as a \emph{bona fide} IC in the usual orbit determination process would be straightforward, since ICs would have a highly hyperbolic orbit, i.e., eccentricities clearly greater than 1. Future advanced sky surveys, particularly the Large Synoptic Survey Telescope (LSST), will be many times more sensitive than past or present observations \citep{LSST}. It is therefore natural to consider whether LSST will detect ICs.  

The motivation for finding ICs is two-fold: discovering an IC would provide new observational opportunities and an IC would be an \emph{in situ} sample of another solar system. Like the discovery of the population of asteroids $\sim$200 years ago and the discovery of the Kuiper belt 20 years ago \citep{1993Natur.362..730J}, the eventual discovery of ICs will open new and unique avenues for exploration that will improve our understanding of the formation and evolution of planetary systems. Photometric, astrometric, and spectroscopic investigations can reveal estimates of the origin, physical, and chemical properties of a piece of another solar system. Even without detailed follow-up observations, the currently unknown frequency of ICs is a useful insight into the efficiency of planet formation in the galaxy. By estimating the frequency at which we expect to observe ICs and then comparing the expected value to the actual observational frequency, we can adjust planet formation models accordingly \citep[e.g.,][]{1990PASP..102..793S}. 

No matter the particulars, discovering ICs or placing upper limits on their frequency would help us to place our solar system in galactic context. However, as a rarely observed population with unique orbital properties, searching for ICs in LSST data will likely require a significant dedicated effort. The value of this effort depends partly on whether the frequency of ICs detected by LSST will have the power to discriminate between different planet formation models. For this reason, we provide a careful assessment of the sensitivity of LSST to different parameters of the IC population.  

The assessments of the frequency of detectable ICs has been highly variable ($\S$\ref{sec:back}). Earlier studies predicted very high numbers of observable ICs, usually by taking what was known from the formation of our solar system, estimating how many comets our solar system ejected into interstellar space, and then multiplying by the number density of stars. For example, \citet{McGlynn89} predicted that the number density of 1 km ICs was approximately $10^{13}$~pc$^{-3}$ and that this implied that several ICs should have already been detected. Recently, a careful assessment of the frequency of ICs by \citet[][hereafter M09]{M09} showed that the actual mass density is orders of magnitude less than these previous estimates, resulting in a number density for ICs larger than 1 km of $10^{5-10}$~pc$^{-3}$. M09 is the first study to self-consistently account for several realistic properties of the IC population by incorporating the stellar mass function, giant planet frequency estimates, solar system minor planet size distributions, and a more recent understanding of the formation of planetary systems. 

M09 considered the detectability of ICs by LSST, providing a clear explanation of why we have not observed any ICs to date. However, their analytical investigation was limited to considering ICs at the distance of Jupiter and beyond. They concluded that LSST would not be able to detect ICs at this distance. However, there are several aspects that may significantly enhance the frequency of detectable ICs closer than Jupiter: gravitational focusing by the Sun would enhance the concentration of ICs within Jupiter's orbit; ICs may brighten by several magnitudes due to outgassing; much more frequent smaller bodies can be seen at closer distances in a magnitude limited survey; etc. 

In order to estimate the realistic detectability of interstellar comets, we have developed a numerical simulation in order to consider all of the factors that play a role in detecting ICs. Our model includes effects from:
\begin{itemize}
    \item increased density due to gravitational focusing (the effect of the Sun altering the trajectory of ICs\index{gravitational focusing}); 
	\item photometric phase functions (the effect of observing ICs at different angles \index{photometric phase angle}); 
	\item comet brightening (accounting for the increase in brightness of comets as they approach the Sun); 
	\item conditions required for observability such as solar elongation (the angular distance between the IC and the Sun) and air mass (e.g., constraints on the altitude of topocentric observations). 
\end{itemize} 
These realistic factors will be discussed in full detail in $\S$\ref{sec:methods}. In addition, our method allows for a determination of many other IC properties relevant to observers, such as typical orbital parameters, rates of motion, and sky distribution. 

It is worth noting that the results of this paper can be divided into two parts. The orbit propagation and astrometry is based on some small assumptions, but is mostly robust. The estimation of the number of ICs that could be detected by LSST, on the other hand, requires several assumptions, in some cases using quantities not known even to within an order of magnitude which we leave as tunable parameters. In this regard, we occasionally neglect effects that would change the highly uncertain results by a factor of $\sim$2. The large uncertainty in our estimates should not be seen as a drawback of the model, but rather a motivation to search for ICs in order to place constraints on their currently unknown properties, with implications for planet formation theory.

\section{Background}
\label{sec:background}

\subsection{History and Results of Other IC Studies}
\label{sec:back}
The currently accepted model of the origin of long-period comets was not always widely accepted. After some initial work by \citet{Opik1932}, the beginnings of the modern model of an isotropic cloud of comets tenuously bound to the solar system emplaced by the planet formation process was originally posed as a ``hypothesis'' by \citet{Oort1950}. At that time, only $\sim$20 long-period comets had well-determined orbits after correcting for planetary perturbations. One major alternative hypothesis was that all (long-period) comets were interstellar, with perhaps an unseen stellar companion to the Sun helping to capture them \citep[e.g.,][]{1982ApJ...255..307V}. The distribution of long-period comets and other small bodies in the solar system now give overwhelming evidence for cometary origins in the Oort cloud, though the origin and history of this cloud is still under discussion \citep[e.g.,][]{Levison2010,2011Icar..215..491K}. 

One of the earliest references to ICs in the context of the Oort cloud origin for long-period comets, is the estimate of \citet{1975AJ.....80..525W} on the frequency of ICs from their non-detection and from the frequency of gamma-ray bursts. As required for an unobserved population, \citet{1975AJ.....80..525W} made various assumptions to estimate the mass density of ICs in the galaxy to be less than $3 \times 10^{-4} M_{\odot}$~pc$^{-3}$. Similarly, \citet{1976Icar...27..123S} estimated an upper limit to the mass density of comets of $\lesssim$$6 \times 10^{-4} M_{\odot}$~pc$^{-3}$ based on the non-detection of ICs up to that point. 

Based on updated estimates of the number of Oort cloud comets, \citet{McGlynn89} use a simple model to estimate that the number of comets is $\sim$10$^{13}$~pc$^{-3}$. If all of these are assumed to have $\sim$1 km in radius, this suggests a mass density of approximately $10^{-5} M_{\odot}$~pc$^{-3}$. These estimates effectively took the number of Oort cloud comets expected from the formation of the solar system ($\sim$10$^{14}$ at the time) and multiplied this by the local density of stars. This is well explained by \citet{1990PASP..102..793S}, who explicitly consider how the frequency of ICs can be used to infer properties of planet formation in the galaxy. 

Upon finding that we have not seen the predicted number of ICs, \citet{McGlynn89} attempt to draw strong conclusions, despite the simplicity of their model for the frequency and detectability of ICs. As a response, \citet{1993A&A...275..298S} suggested that the stellar density used by \citet{McGlynn89} was an order of magnitude too high, bringing the expected frequency of detectable ICs down enough that they were no longer missing from the observations. 

Using a much more detailed model and a survey simulator,  \citet{Francis05} placed a limit of $3 \times 10^{12}$ ICs per cubic parsec based on a non-detection of ICs in the LINEAR survey. If these are all assumed to be $\sim$1 km in radius, this suggests a mass density of $3 \times 10^{-6} M_{\odot}$~pc$^{-3}$, two orders of magnitude lower than the earliest estimates. 
\citet{2004DPS....36.4008M} mention a study of the Spaceguard survey which found a 97\% upper limit on 1 km ICs of 10$^{14}$ ICs per cubic parsec, but included a power law distribution (corresponding to $q_1 = q_2 = 3.5$ in our nomenclature below). 

The most recent study \citep{2014acm..conf..149E} used Pan-STARRS 1 data, a careful analysis of detectability, a power-law distribution (with $q_1 = q_2 = 3.5$), and included the possibility of cometary activity. This most sophisticated method has similarities with our methodology described below. The 90\% confidence upper limit on the number density of $\gtrsim$1 km ICs per cubic parsec was $4.7 \times 10^{14}$ for inactive comets (``asteroids'' in our nomenclature) and $1.6 \times 10^{13}$ for active comets. This is not as strong a limit as the \citet{Francis05} result, but is the result of a more detailed assessment. It is also able to roughly rule out the \citet{McGlynn89} estimate. 

M09 was the first theoretical estimate to use a more modern understanding of planet formation, accurate distributions for the frequency of stars of different masses in the Galaxy, and incorporating an IC size distribution to determine a much more realistic estimate for the frequency of ICs. Although their model makes many assumptions about the unknown properties of planet formation and IC properties, these are mostly included in the form of tunable parameters. M09 estimate the mass density of ICs to be $\sim$$2 \times 10^{-7} M_{\odot}$~pc$^{-3}$, which, when considering a size distribution for ICs, yields a number density of ICs larger than 1 km of $10^{5-10}$~pc$^{-3}$. This is between 3 and 8 orders of magnitude below the \citet{McGlynn89} estimate and the \citet{2014acm..conf..149E} upper observational limit. 

Following \citet{1986ApJ...302..462A}, \citet{JuraM11} has also recently shown that the space density of ICs must be less than predicted by the earlier optimistic assessments by studying the chemical composition of unpolluted white dwarfs (which would have retained signatures of accreted ICs). They place an upper bound on the space density of ICs in agreement with M09 and far less than \citet{McGlynn89}. \citet{2012MNRAS.421.1315Z} predict a similar frequency of ICs near the galactic center as a possible source of Sgr A* flares. 

Besides illustrating the decline in IC frequency estimated over time, these studies show that simplifying assumptions and unknown properties can yield IC detectability estimates that span many orders of magnitude. Even intercomparing the results of these studies requires assumptions about the size distribution or typical size of detectable comet nuclei and the mass-radius-brightness relation of comet nuclei, neither of which are known very well. As in M09, we have tried to use tunable parameters to understand the importance of various assumptions. By using the most up-to-date theory from M09 with a model that incorporates realistic estimates of detectability, we have produced the most sophisticated estimate of the frequency of detectable of ICs to date. 

\subsection{A Candidate Interstellar Comet?}
\label{sec:candidate}

\citet{2013MNRAS.435..440K} and \citet{2015MNRAS.448..588D} find that comet C/2007 W1 (Boattini) is a strong candidate interstellar comet. While some comets appear slightly hyperbolic due to interactions with the giant planets and/or non-gravitational forces, an analysis of these effects show that, in this case, they are much too small to explain C/2007 W1's orbit with an initial semi-major axis of -23000 AU ($1/a = -42.75 \pm 2.34 \times 10^{-6}$ AU), perihelion of 0.83 AU, and inclination of 10$^{\circ}$ \citep{2015MNRAS.448..588D}. Over 1000 observations of C/2007 W1 were obtained over 13 months, sufficient to produce a high quality orbit, as reflected on the Minor Planet Center, which has similar orbital estimates. Another candidate interstellar comet, C/1853 E1 (Secchi), relies on century-old astrometry and can be probably attributed to systematic errors \citep{2012AN....333..118B}. 

C/2007 W1 was observed spectroscopically and found to be an unusual comet chemically, as might be expected for an IC \citep{2011Icar..216..227V}, although it wasn't completely out of the range of known comets. It was also the source of a somewhat unusual meteor shower \citep{2011MNRAS.414..668W}.

Converting the excess energy in C/2007 W1 from \citet{2015MNRAS.448..588D} to a velocity at infinity gives $v_{\infty} \simeq 0.27 \pm 0.01$ km s$^{-1}$, much smaller than the expected $\sim$20 km s$^{-1}$ for an IC. Along the same lines, C/2007 W1 has a post-perihelion orbit that is bound to the solar system (semi-major axis of 1800 AU, perihelion distance of 0.85 AU) which seems unusual for an IC, though this isn't an entirely separate argument, since the capture is highly enhanced due to the low $v_{\infty}$. In our simulations, we do not find a particularly enhanced preference for detecting low $v_{\infty}$ ICs, so this low relative velocity is seemingly an argument against the interstellar origin of C/2007 W1.

Altogether, the orbital and chemical evidence for the interstellar nature of C/2007 W1 is highly suggestive, but not conclusive. We, therefore, proceed without including this comet in our analysis.

\subsection{Properties of the Source Region of Interstellar Comets}

During the lifetime of LSST (10 years), even ICs with unusually high (100 km/s) relative velocities to the Sun will only traverse approximately 200 AU or $\sim$0.001 parsecs. Therefore, the dominant medium of ICs observable in the next several decades is the region of space extremely close to the Sun, in a galactic sense, and dominated by the so-called Circumheliospheric Interstellar Medium (CHISM or CISM), sometimes called the Very Local Interstellar Medium (VLISM). This region is much smaller than the Local Interstellar Cloud ($\sim$10 pc) or Local Bubble ($\sim$100 pc), and not even much larger than the heliosphere ($\sim$100 AU). Indeed, any IC passing by Earth within the next decades is currently residing within the inner Oort cloud, far closer than the aphelion of Sedna and other Kuiper belt objects \citep{2004ApJ...617..645B}. Here we briefly review the known properties of the IC source region. 
In situ samples of the CHISM as it flows into the inner solar system are taken by observing interstellar neutral atoms, interstellar pickup ions, interstellar dust grains ($\lesssim$1$\mu$m, entrained in the local gas flow), and interstellar micrometeorites ($\gtrsim$10$\mu$m and decoupled from the gas). Observations of the CHISM over the last 4 decades have not shown any significant evolution, suggesting that the properties are homogeneous on the scales relevant to near-future IC detections (Frisch et al. 2009). The mass ratio of gas to dust in the CHISM is thought to be $\sim$100 (Frisch et al. 1999) and the dust mass density is about $\sim$100 times bigger than the mass in ICs estimated by M09 \citep{2000JGR...10510343L}. 

Dust smaller than $\sim$10 microns is observed in situ by dust detectors on NASA spacecraft (e.g., Ulysses, Cassini, Helios, and Stardust) in deep space. The observations show that these grains are fully entrained in the local interstellar flow, as expected since they are small enough to couple to the gas, even at the low CHISM gas densities (Frisch et al. 2011). The NASA Stardust mission returned samples from its Interstellar Dust Collector, which was designed to capture interstellar grains directly from this local flow \citep{2014M&PS...49.1720W}. 

\label{sourceregion}

On the other hand, the trajectories of interstellar dust larger than $\sim$10 microns is dominated by solar gravity in the region of the Sun and can travel unperturbed for hundreds of parsecs from its source region where they are detected as Earth-impacting micrometeorites by ground-based optical and radar observations \citep{2000JGR...10510353B,2004ApJ...600..804M,2012ApJ...745..161M}. Very little is known about the frequency or properties of larger particles \citep{2009ApJ...702L..77S}. Observations of these interstellar micrometeorites have detected a broad ``background'' source, but also a significant discrete source that may be associated with debris-disk and planet hosting star $\beta$ Pictoris \citep{2000JGR...10510353B,2010Sci...329...57L}. \citet{2004ApJ...600..804M} suggest that large grains may indeed be from discrete sources. However, there is a chance that these interstellar meteoroids are contaminated by solar system particles that have reached hyperbolic velocities due to planetary encounters \citep{2014Icar..242..112W}. Generally, these have lower velocities than the expected incoming stream of interstellar micrometeorites, but it does present a source of confusion. See \citet{2014Icar..242..112W}, \citet{2014M&PS...49...63H} and references therein for additional discussion on interstellar micrometeorites and meteorites. 

Direct observation of ICs with LSST or other optical surveys probes the local IC population well above 1-meter sizes. M09 (and our own simulations) found that the most detectable objects will be the smallest ICs which, though intrinsically fainter, typically pass much closer than the larger ICs which more than compensates for their smaller area. Large ($\gtrsim$100 km) ICs are so rare, that they are very unlikely to be seen. 

We are not considering objects in the $\gtrsim$10000 km size regime, as these unbound planet-sized objects (variously called free-floating planets, rogue planets, interstellar planets, or nomads) can have intrinsic luminosity and have been studied elsewhere \citep[e.g.,][]{2011ApJS..197...19K,2012MNRAS.423.1856S}. 

\section{Methods}
\label{sec:methods}

We estimate the number of ICs that are detectable with optical surveys. To do so, we simulate the detectability of billions of ICs numerically. Starting with estimates for the physical and orbital properties of the IC population, we calculate their brightness as a function of time based on empirical relations developed for comets and asteroids. Potentially detectable ICs are investigated in more detail to determine their visibility to LSST. Throughout we keep track of tunable parameters, noting that the unknown frequency of ICs means that our final results can vary by at least an order of magnitude. Still, we have attempted to be accurate in our modeling, in order to understand the importance of various effects.\footnote{The code is available upon request.}

\subsection{Physical and Orbital Properties}
\label{sec:init}

\subsubsection{Initial Position and Velocity}
Our main mode of simulation assumes that ICs have been well-mixed and form an isotropic population. Therefore, we simulate a large cube centered on the Sun within which the initial positions of ICs are randomly and isotropically generated (but see $\S$\ref{discrete} below). The cube is chosen to be large enough (1000 AU) that our simulation is insensitive to edge effects. 

In keeping with our isotropic assumption, for each IC we choose a single, initial, randomly-oriented velocity ($v_0$). To track the importance of the velocity in IC detectability, we choose to do many independent runs with constant initial velocities rather than a velocity dispersion in a single run, but this does not affect our conclusions. These simulations confirmed that slower ICs were more concentrated towards the Earth and Sun (due to gravitational focusing), with about twice as many comets with $v_0$ of 5~km~s$^{-1}$ coming within $\sim$5 AU than comets with $v_0=30$~km~s$^{-1}$. Over the age of the universe, even high-velocity ICs will only typically intersect their own mass if they are smaller than $\sim$1 cm, suggesting that larger particles are completely decoupled from gas in the interstellar medium. These ICs therefore keep their original velocity with which they were originally ejected and are stirred collisionlessly in the gravitational potential of the galaxy, like stars, and should have a similar velocity dispersion of tens of km s$^{-1}$. 

The Sun is not fixed with respect to the galactic reference frame of ICs. The Sun and its vicinity are rotating together around the center of the galaxy. To isolate the local relative motions relevant here, a construct called the Local Standard of Rest (LSR) is used, which is a coordinate system moving around the galaxy in a circular orbit at $\sim$220 km/s. The Sun's ``peculiar'' motion relative to the LSR is not known precisely; we use a recent proposed LSR velocity for the Sun from \citet{SBD10}: (($U_{\odot}, V_{\odot}, W_{\odot})_{LSR} \simeq (11.1, 12.2, 7.3)$ km s$^{-1}$). To keep the Sun fixed in the center of the simulation, we subtract this velocity from each IC, which preserves the notion that the Sun is flying through an isotropic background of ICs, though they also have their own isotropic velocity dispersion \citep[e.g.,][]{1975AJ.....80..525W}. We neglect the tiny Coriolis force due to the rotating reference frame. 

The Sun's peculiar velocity relative to the LSR is based on the mean motion of main sequence stars assumed to be moving together around the galaxy; the distances to these stars is much greater than the current source region of ICs, but the velocity of the Sun relative to the local interstellar wind or flow is quite similar. Though there are slight differences (usually within the error bars), the entrained dust, the neutral helium, and the cluster of local interstellar clouds are all traveling at nearly the same velocity and uniform direction relative to the Sun (Frisch et al. 2011) which is slightly faster (by 6.6 km/s) and $\sim$40 degrees offset from the direction of the Sun's peculiar motion relative to the LSR. There is some evidence that the flow direction is even changing with time \citep{2012arXiv1210.7214F}. However, our simulations show that the direction and magnitude of the solar velocity does not affect the detectability of ICs and we use the LSR-relative motion for simplicity. 

\subsubsection{Orbital Motion and Gravitational Focusing}

Since they are not bound to the Sun, ICs follow hyperbolic orbits. Using the aforementioned initial heliocentric position and velocity, the entire orbital path is determined using the standard equations and osculating orbital elements for hyperbolic orbits in the two-body problem. We do not integrate the orbits of the ICs and therefore do not account for perturbations by the planets \citep{1986AJ.....92..171T} or non-gravitational forces \citep[due to outgassing, e.g.,][]{2011AJ....142...81A}, since these are both negligibly small for our statistical purposes. By calculating their hyperbolic orbits explicitly, our simulation automatically accounts for gravitational focusing: the excess of ICs that will pass near the Sun due to its gravitational influence. This is not present in the initial placement of the ICs, but as our simulations run forwards and backwards in time for thousands of years, the initial isotropic placement is acceptable. 

We also calculate the motion of the Earth (with an arbitrary phase). From this we can determine at any time $\Delta_{earth}$, the distance from the IC to the Earth, and $\Delta_{sun}$ is the distance from the IC to the Sun. Furthermore, we can calculate the astrometric position (Right Ascension and Declination) of each IC as a function of time, as discussed below.

\subsubsection{Broken Power Law Size Distribution and Mass Density}

The number and size of the ICs that are initially placed in the simulation cube is determined by using a broken power law size distribution and an overall mass density ($m_{total}$). The size distribution defines the number of ICs we expect to exist for a given radius. Throughout this paper the ``size'' of an IC refers to the radius of the comet nucleus. The mass density is the amount of mass we expect exists in a given volume. The size distribution and mass density are combined by assuming that all ICs have the same density (nominally 0.5 g cm$^{-3}$). As a result we can calculate the number of ICs that would exist in a given volume along with their respective radii. 

Our nominal simulation adopts the mass density of M09, with the total mass of ICs of $m_{total} = 2.2 \times 10^{-7} M_{\odot}$ pc$^{-3} = 4.5 \times 10^{26}$ g pc$^{-3} = 5.1 \times 10^{10}$ g AU$^{-3}$. If the mass in a cubic AU were concentrated into a single object with the density of water, it would only be 23 meters in radius. This is ridiculously sparse, which explains why ICs have evaded detection thus far, even though their $\sim$4 AU/year motion relative to the Sun is constantly replenishing the possibility of detection. 

We note that, due to the uncertain nature of planet formation and the creation and ejection of ICs, the estimate of M09 could be substantially in error. Several effects could increase the mass density of ICs each by a factor of a few: an updated stellar density \citep{2012MNRAS.425.1445G}, Oort cloud stripping from galactic tides \citep{2013MNRAS.430..403V}, ejection of Oort clouds due to the death of stars \citep{2011MNRAS.417.2104V,2012MNRAS.421.2969V}, and many other possible effects. Since M09, the \emph{Kepler Space Telescope} has also discovered entirely new classes of planetary systems, calling into question aspects of planet formation theory used to justify the estimates of M09. 

\begin{figure}
 \centering
\includegraphics{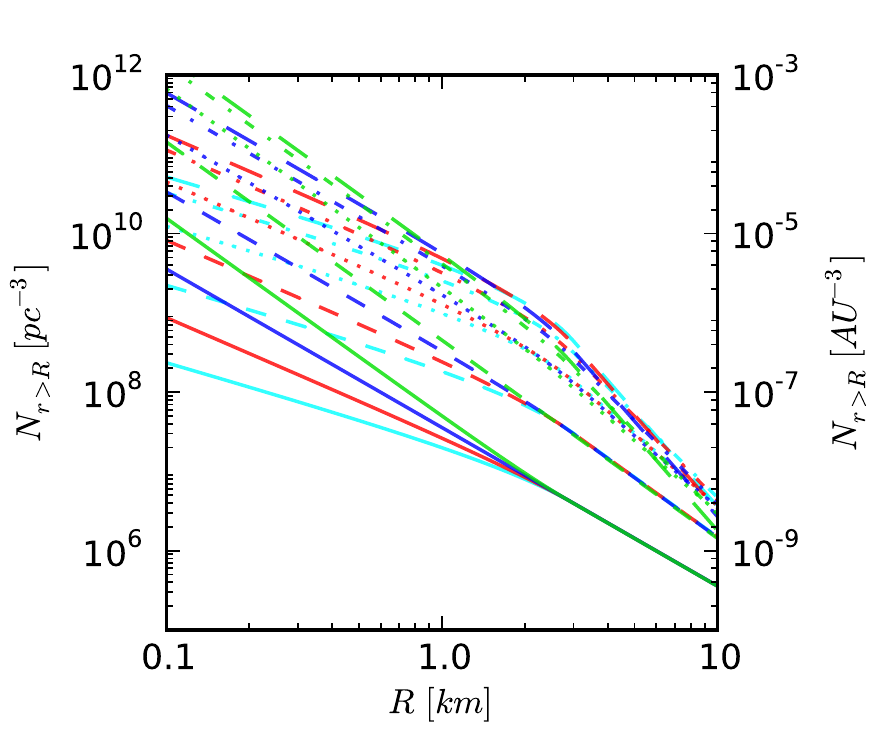}
 \caption[Number density plot from Moro-Mart{\'{\i}}n.]{Interstellar comet (IC) cumulative number density per cubic parsec and per cubic AU as a function of different power law parameters following M09. The M09 IC mass density of $m_{total} = 2.2 \times 10^{-7} M_{\odot}$ pc$^{-3}$ is used. We use a broken power law; shown are the distributions with a break radius at 3 km. This figure also corrects a small error in Figure 1 of M09. The colors and line types correspond to $q_1 = 2.0$ (light blue, lowest), 2.5 (red), 3.0 (blue), and 3.5 (green, highest); $q_2 = 3$ (solid), 3.5 (dashed), 4 (dotted), 4.5 (dash-dotted), and 5 (long dashed), where $q_1$ is the differential size distribution index for objects below the break radius and $q_2$ is the same for objects above the break radius. Notice the wide variety of number densities at the smallest sizes which translates to significant uncertainty in the detectability of ICs. \label{fig:m1correction} \index{broken power law}}
\end{figure}

Our simulations use the size distribution prescriptions suggested by M09. In particular, the size distribution is a broken power law, with a variable break radius $(r_b)$ and different size distribution slopes on each side of the break. Following M09, we define differential size distribution slopes $q_1$ and $q_2$, such that $n(r) \propto r^{-q_1}$ if $r < r_b$ and $n(r) \propto r^{-q_2}$ if $r > r_b$. We place practical limits on the minimum and maximum radii in our simulations (see below). The number densities as a function of different $q_1$ and $q_2$ values for $r_b=3$ km are shown in Figure \ref{fig:m1correction}. By detecting serendipitous KBO occultations, \citet{2012arXiv1210.8155S} have estimated that $q_1 \approx 2.8 \pm 0.1$, while \citet{2009AJ....137...72F} and \citet{2009ApJ...696...91F} suggest that $q_2 \approx 4.5$ with a break radius of $r_b \approx 75$ km in the Kuiper Belt \citep[see also limits from the lack of detection of KBOs by WMAP,][]{2011ApJ...736..122I}. See M09 for more discussion on the possible size distribution relevant for ICs; here we simulate several different possibilities. 

Our ``nominal'' model uses typically assumed values for the variables that describe the properties of ICs \emph{except for the size distribution, where we use a very optimistic case}. It is important to remember throughout that the observational and theoretical estimates of parameters in our ``nominal'' model are sometimes controversial and often with significant uncertainty. This reemphasizes the importance of leaving many variables as free parameters. 

Finally, we note here that there was a small error in the number density equation as derived by M09. Their number density equation (Equation 5 in M09) is only valid for radii less than the break radius because of the limits of integration used during its derivation (A. Moro-Mart{\'{\i}}n, personal communication). A piecewise equation is needed to correctly define the number density for radii less than and greater than the break radius. We have made this correction and reproduced a plot of the number density of the ICs in Figure \ref{fig:m1correction}, showing the correction to their Figure 1 for their nominal mass density. Since the number density and observability of ICs is completely dominated by the small objects, their errors above the break radius have no consequence for the results reported in M09. 

\subsection{Calculating IC Brightness}
\label{sec:magnitude}

\subsubsection{Asteroid Case}

Based on the above calculations, we now have the position of the IC, Sun, and the Earth at any time. Determining the brightness is based initially on the standard apparent magnitude equation for asteroids\citep{1989aste.conf..524B}: 
\begin{equation}
\label{eq:dismag}
V = H + 2.5  \left[ \log_{10} (\Delta_{sun}^2) + \log_{10} (\Delta_{earth}^2) \right] - 2.5 \log_{10}(\gamma)
\end{equation}
where $H$ is the (asteroid) absolute magnitude, related to the intrinsic brightness of the asteroid independent of the observing geometry and $\gamma$ is the correction for photometric phase described below. Specifically, $H$ is the brightness an asteroid would have if it were 1 AU from the Sun and from the observer and observed with the Sun-asteroid-observer angle of zero (an observing geometry that is only possible if the observer is at the position of the Sun). The absolute magnitude is intrinsic to the body and determined by radius and albedo in the asteroid case. We use the standard definition
\begin{equation}
\label{eq:absmagas}
	 H = \frac{\log_{10}\left(2r\right)  + 0.5 \log_{10}(p) - 3.1236}{-0.2}
\end{equation}
where $r$ is the radius in km and $p$ is the albedo. Following M09, our nominal assumed albedo is 0.06, but this is actually only applied to asteroids (the connection between comet brightness and nucleus radius is handled in a different way).

The photometric phase angle is the Sun-IC-Earth angle with the IC at the vertex. To encapsulate the effects of diminished reflection and non-uniform surface scattering and non-zero phase angles, we follow standard methods of using a phase function $\gamma$ to adjust the brightness of the ICs based on its phase angle. The phase function we use is the standard function from \citet{Muinonen}:
\begin{equation}
\label{eq:phase}
\gamma = (1 - G)\Phi_1(\theta) + G\Phi_2(\theta),
\end{equation}
where $G$ is a slope parameter, $\Phi_1$ and $\Phi_2$ are basis functions for the phase curve (Equation 6 from \citet{Muinonen}), and $\theta$ is the phase angle. The value of $G$ controls how steep the phase curve is; values close to 0 indicate a steep curve and values close to 1 indicate a shallow curve. We use a steep curve with $G=0.15$, the standard value used for objects with unmeasured phase curves. For the most part, the photometric phase angle correction is not significant unless observations are taken at large angles (i.e., far from opposition), when it can drop the brightness of an object by several magnitudes (mostly because of the smaller ``day'' side). This can be relevant for isotropically distributed ICs which may be seen in non-standard geometries.

Throughout, we refer to objects that follow this photometric prescription as ``asteroids'' although in practice they may be dormant or inactive comet nuclei. 

\subsubsection{Comet Brightening}

As comets approach the Sun they can become active and have a significant increase in intrinsic brightness. There are two aspects of comet brightness to consider: how the radius of the comet nucleus relates to its brightness and how the intrinsic brightness grows in time as the comet approaches the Sun and becomes more active. In both cases, we use the standard empirical methods to determine the brightness of ICs.

To relate the radius of comet nuclei to their brightness, we use the comet absolute magnitude, which we call $H_c$ in order to distinguish it from the asteroid absolute magnitude, which is defined differently. The relation to determine $H_c$ for comets is
\begin{equation}
\label{eq:absmag}
	 H_c = \frac{\log_{10}\left(2r\right) - b_2}{b_1},
\end{equation}
where $r$ is again the radius of the comet nucleus in km and $b_1$ and $b_2$ are empirical parameters determined from observational data. These terms absorb the albedo term seen in Eq. (\ref{eq:absmagas}); the asteroid case is equivalent to $b_1$=-0.20 and $b_2 = 3.1236 - 0.5 \log_{10} (p) = 3.73$ using $p=0.06$. Several different values of $b_1$ and $b_2$ have been estimated as noted in Table \ref{tbl:cb}, therefore, we leave these as free parameters of our model. The nominal model uses $b_1 = -0.13$ and $b_2=1.20$, the most recent estimates from \citet{Sosa11}. 
\begin{table}
	\begin{center}
	\begin{tabular}{l|ll}
		 Source &  $b_1$ & $b_2$ \\
		\hline
		Kresak 1978 & -0.20 & 2.10 \\
		Bailey \& Stagg 1988 & -0.17 & 1.90 \\
		Weissman 1996 & -0.13 & 1.86 \\
		Sosa \& Fernandez 2011 & -0.13 & 1.20 \\
	        asteroid with albedo 0.06 & -0.20 & 3.73 \\
	\end{tabular}
	\caption[]{List of the empirical comet parameters that relate comet nucleus radius with intrinsic brightness using Equation \ref{eq:absmag}. The influence on comet brightness from these parameters is given in Figure \ref{fig:cb_D}.\label{tbl:cb}}
	
	\end{center}
	\nocite{Kresak78,Bailey88,Weissman96,Sosa11}
\end{table}

\begin{figure}
 \centering
 \includegraphics{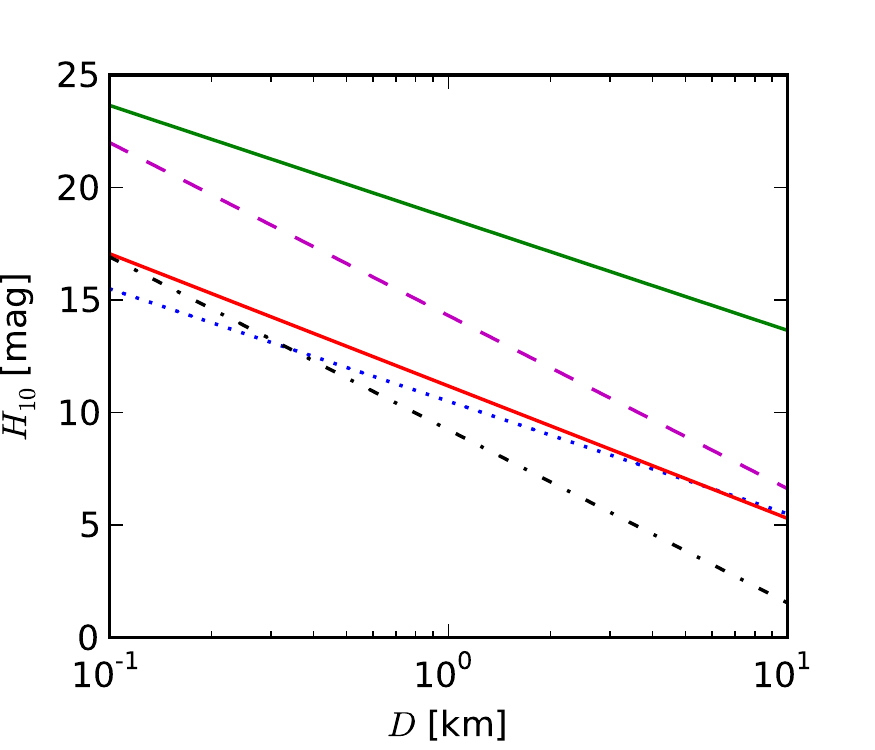}
 \caption[Plot of $H_c$ for the different sources of $b_1$ and $b_2$.]{Plot of $H_c$ for the different empirical comet nucleus size-brightness parameters $b_1$, which controls the slope, and $b_2$, which controls the offset (Table \ref{tbl:cb} and Equation \ref{eq:absmag}). The lines correspond to the different sources as follows: 6\% albedo asteroid (green, solid, upper), Kresak 1978 (blue, dotted), Bailey \& Stagg 1988 (red, solid, lower), Weissman 1996 (magenta, dashed), Sosa \& Fernandez 2011 (black, dash dotted). These illustrate how comets are much brighter than equivalent-size asteroids and the significant uncertainty involved in the comet size-brightness relations.\label{fig:cb_D}}

\end{figure}

After determining the absolute magnitude of the comets based on Eq.~(\ref{eq:absmag}), an additional term is needed to model how the intrinsic brightness of comets grows due to increased radiation from the Sun near perihelion. Following the standard in the comet community, we let the brightness vary as $1/\Delta_{sun}^n$ where $n$ is an adjustable parameter called the photometric index \citep[e.g.,][]{Sosa11}. When $n=2$, this reduces to the case without comet brightening. Comets are typically modeled with two different values of $n$, one for the pre-perihelion approach and another for the post-perihelion orbit. We use a pre-perihelion $n$ of 5.0 and a post-perihelion $n$ of 3.5, standard for long period comets thought to originate from the Oort cloud, which are the best analog for ICs in this regard \citep{Francis05}. Large values of $n$ correspond to steeper brightening functions. The photometric index is based on empirical observations of comets that are generally based on detections within $\lesssim$5 AU and should be determined in conjunction with $H_c$, $b_1$, and $b_2$ \citep[e.g.,][]{Sosa11}. In our analysis, we allow $H_c$ to be defined by the assumed nuclear radius ($r$) using Equation \ref{eq:absmag} independently of any correlation that may exist between $n$, $b_1$, and $b_2$. This affects the direct comparison to comet studies in the solar system, but is still a reasonable approximation in the face of other systematic errors and uncertainties in comet parameters. We also note here that the common comet absolute magnitude $H_{10}$ is equivalent to $H_c$ under the assumption of a photometric index of $n=4$, e.g., a brightness that depends on the Sun-comet distance to the fourth power. 

\begin{figure}
	\centering
	\includegraphics{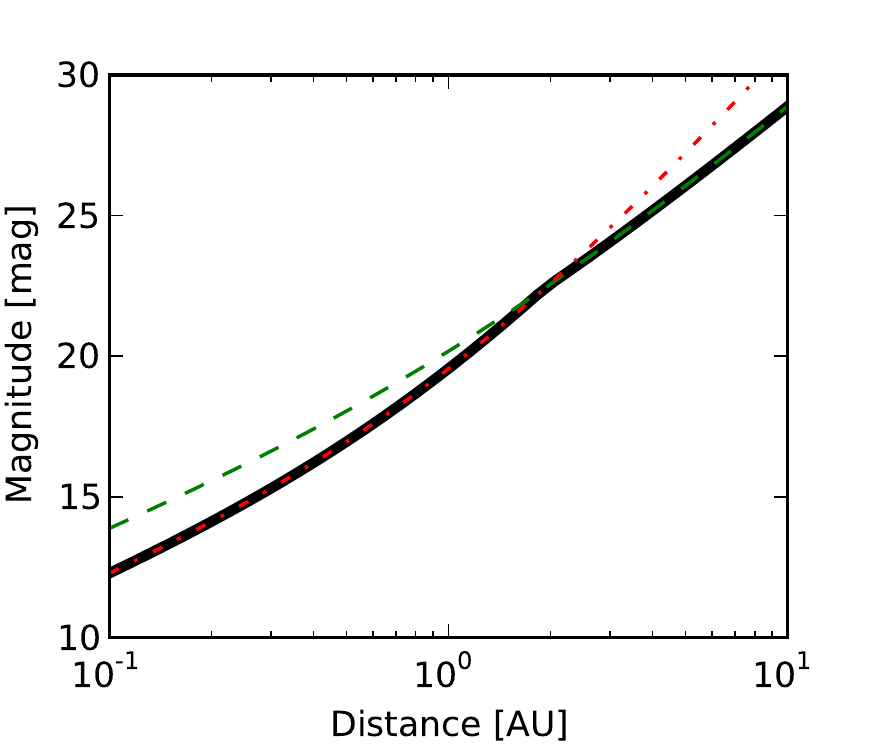}
	\caption[Asteroid magnitude vs. comet magnitude at opposition]{Asteroid magnitude vs. comet magnitude at opposition for an object with 1 km radius. Distance refers to the Earth-IC distance ($\Delta_{earth}$) and the Sun-IC distance is $\Delta_{sun} = \Delta_{earth} + 1$ in this illustration. The red dashed-dotted line represents the magnitude equation for a comet with a photometric index of $n=5$. The green dashed line is the magnitude of an asteroid. Since the comet brightening parameter inappropriately penalizes the brightness at large distances, when modeling the brightness of an IC, we use whichever is brighter (solid black line). Both comets and asteroids are far brighter when they are close to the Earth/Sun, but cometary activity can increase the brightness even further.\label{fig:ciccbmags}}
	
\end{figure}

At distances far beyond Jupiter, the photometric index strongly penalizes the comet magnitude in an unrealistic way. Although it does not affect our results, we avoid this by requiring that the magnitude of an IC be always less than (e.g., brighter than) the magnitude of an interstellar asteroid with 6\% albedo. Using this piecewise definition we ensure that a comet is never fainter than an asteroid of equivalent properties. A typical example of the brightness of a comet and asteroid in our model is shown in Figure \ref{fig:ciccbmags}. 

The full magnitude equation for comet brightening then becomes:
\begin{equation}
\label{eq:cbmag}
	V = H_c + 2.5  \left[ \frac{n}{2} \ log_{10}(\Delta_{sun}^2) + \log_{10} (\Delta_{earth}^2)\right] - 2.5 \log_{10}(\gamma)
\end{equation}
which includes the most relevant effects. We do not consider intrinsic brightness variations due to rotational modulation or outbursts and we assume that any extended brightness due to a coma is smaller than the photometric aperture. 

\subsection{Criteria for Determining Detectability}

Using the above methods, we can take a specific population of ICs and determine their brightness at any time. The vast majority of ICs in the initial simulation are completely undetectable due to the large region of space modeled (1000 AU cube). To produce results that are robust from small number statistics, we use a Monte Carlo approach of simulating billions of ICs in order to ensure that a significant number (usually $\gtrsim$100) of detectable ICs are generated. This necessitates the use of renormalization between simulated ICs and the actual proposed population of ICs, e.g., the orbital path of each simulated IC can represent a variable number of ICs that would actually be present, given the parameters of the simulation. We also calculate IC observations over a few thousand years and then average the non-transient portions of these observations in order to determine the rate of detectable ICs, even when this rate is very small. 

The first step in narrowing down the large number of simulated ICs into those that are potentially detectable is an optimization routine that determines the minimum magnitude of a particular IC. We begin this routine with an initial guess that maximum brightness is at perihelion. In reality, there can be multiple maxima in brightness due to the orbital motion of the Earth. However, we determined that ICs that had a minimum magnitude of fainter than 28 in this first step will never be detectable by LSST. 

The second step takes these potentially detectable ICs and computes their observational parameters on a much finer time grid that covers the possible range of observable times, based on the minimum magnitude found earlier. Every few hours, the brightness, solar elongation, airmass, and other parameters are determined. We consider an IC ``detectable'' or ``observable'' if it meets the following criteria during at least one timestep:
\begin{enumerate}
\item IC magnitude less than the limiting magnitude
\item Solar elevation less than -18$^{\circ}$ (end or beginning of astronomical twilight)
\item IC airmass less than 2 as observed from Cerro Pachon (future site of LSST)
\end{enumerate}
Note that the latter two effects automatically require the solar elongation to be greater than 48$^{\circ}$. This is important since comet brightening is extremely and unrealistically enhanced when the Sun-IC distance is very small. We do not consider the effects of the Moon, the specific observing cadence, downtime, imaging fill factor, etc. In practice, we find that detectable ICs are often detectable over several days or weeks, so it is not likely that LSST will miss a substantial fraction of these objects due to these effects, as long as the above detectability criteria are met. We discuss the possible effects of trailing below, but otherwise assume that the survey is 100\% efficient up to the limiting magnitude for simplicity. Note that the meaning of limiting magnitude in surveys is a 50\% recovery rate, while our usage of the term implies a 100\% recovery rate.

\section{Results}
\label{sec:results}

Using our model, we calculate the results of the number of observable ICs detectable by LSST (or other surveys) for different choices of the input parameters. This is summarized in one number, $N_{LSST}$, the number of ICs expected to be realistically detectable over the LSST's 10-year lifespan as a function of limiting magnitude. LSST plans to have a limiting magnitude of about $V\approx24.5$, which is the nominal value we use throughout the discussion. Due to larger uncertainties in other parameters, we do not consider the effects of color or specific filter choices. 

\begin{table*}
    \begin{center}
	\begin{tabular}{l|l|l}
	Parameter & Description & Nominal Value \\
	\hline
 	$m_{total}$ & mass density of ICs & $4.5\times10^{26}$~g~pc$^{-3}$\\
	$q_1$ & slope of the differential IC size distribution when $r<r_b$ & 3.5\\
	$q_2$ & slope of the differential IC size distribution when $r>r_b$ & 5.0\\
	$r_{b}$ & the break radius of the IC number density & 3~km \\
	$r_{min}$ & minimum radius of detectable ICs & 0.1~km \\
	$\rho$ & bulk density of IC nuclei & 0.5~g~cm$^{-3}$\\
	$n_{pre}$ & pre-perihelion photometric index & 5.0\\
	$n_{post}$ & post-perihelion photometric index\index{photometric index} & 3.5\\
	$b_1$ & comet absolute magnitude parameter & -0.13\\
	$b_2$ & comet absolute magnitude parameter & 1.2\\
	$v_0$ & velocity dispersion of ICs & 30~km~s$^{-1}$ \\
	$G$ & phase function steepness & 0.15\\
	$p$ & albedo of asteroid & 0.06 \\
	$m$ & limiting magnitude & 24.5\\
	$\Delta_{earth}$ & minimum geocentric distance allowed & 0 (AU)\\
	\end{tabular}
	\caption{List of input parameters for the simulation and their nominal values.\label{tbl:params}}
	
	\end{center}
\end{table*}

\begin{table*}
    \begin{center}
    \tiny
	\begin{tabular}{l|l|l|llllll}
    
		 \multirow{2}{*}{\#} & \multirow{2}{*}{Label} &  \multirow{2}{*}{Difference from nominal values} & & & $N_{LSST}$ & & \\
    	 & & & 20.5 & 21.75 & 23 & 24.5 & 25.5 & 26.75\\
    	\hline
		1 & Nominal & none & 0.15 & 0.23 & 0.35 & 0.57 & 0.85 & 1.2\\

		2 & Nominal without gravity & no gravity & \nodata & \nodata & \nodata & 0.51 & \nodata & \nodata \\

		3 & Nominal ignoring the phase function & no phase function & \nodata & \nodata & \nodata & 0.95 & \nodata & \nodata\\

		4 & Nominal using the comet density from M09 & $\rho=1.5$~g~cm$^{-3}$, & \nodata & \nodata & \nodata & 0.20 & \nodata & \nodata \\

		5 & Nominal excluding comets within 5 AU & $\Delta_{earth} \geq 5$ & \nodata & \nodata & \nodata & 0.041 & \nodata & \nodata\\

		\multirow{3}{*}{6} & \multirow{3}{*}{Moro-Mart{\'{\i}}n} & $\rho=1.5$~g~cm$^{-3}$, $n_{pre}=2$, $n_{post}=2$, & \multirow{3}{*}{\nodata} & \multirow{3}{*}{\nodata} & \multirow{3}{*}{\nodata} &  \multirow{3}{*}{0.0023} & \multirow{3}{*}{\nodata} & \multirow{3}{*}{\nodata}\\
		& &  $b_1=-0.20$, $b_2=3.1236$, $\Delta_{earth} \geq 5$, & & & & & & \\
         & &  $p=0.06$,  no gravity, no phase function & & & & & & \\

		7 & McGlynn and Chapman mass density & $m_{total}=4.5\times10^{30}$~g~pc$^{-3}$ & \nodata & \nodata & \nodata & 5900 & \nodata & \nodata\\

		8 & minimum size of 0.2 km & $r_{min}=0.2$ & \nodata & \nodata & \nodata & 0.39 & \nodata & \nodata\\

		9 & minimum size of 0.5 km & $r_{min}=0.5$ & \nodata & \nodata & \nodata & 0.22 & \nodata & \nodata\\

		10 & minimum size of 1 km & $r_{min}=1$ & \nodata & \nodata & \nodata & 0.14 & \nodata & \nodata\\

		11 & realistic case & $q_1 = 2.92$, $r_{min}=1$ km & \nodata & \nodata & \nodata & 0.0012 & \nodata & \nodata\\

		\multirow{2}{*}{12} & \multirow{2}{*}{Asteroid} & $r_{min}=0.01$km, $n_{pre}=2$, $n_{post}=2$, & \multirow{2}{*}{\nodata} & \multirow{2}{*}{\nodata} & \multirow{2}{*}{\nodata} & \multirow{2}{*}{0.94} & \multirow{2}{*}{2.9} & \multirow{2}{*}{6.8}\\
		& &  $b_1=-0.20$, $b_2=3.1236$, $p=0.06$ & & & & & & \\

		13 & slow $v_0$ & $v_0=5$~km~s$^{-1}$ & 0.15 & 0.23 & 0.35 & 0.64 & 0.86 & 1.2\\

		14 & Kresak comet parameters & $b_1=-0.20$, $b_2=2.10$ & 0.21 & 0.33 & 0.55 & 1.0 & 1.42 & 2.1\\

		15 & Bailey \& Stagg comet parameters & $b_1=-0.17$, $b_2=1.90$ & 0.11 & 0.19 & 0.33 & 0.58 & 0.89 & 1.39\\

		16 & Weissman comet parameters & $b_1=-0.13$, $b_2=1.86$ & 0.0045 & 0.023 & 0.079 & 0.20 & 0.37 & 0.76\\

		17 & small number density & $q_1=2.0$, $q_2=3.0$ & 0.00006 & 0.00010 & 0.00017 & 0.00029 & 0.00041 & 0.00060\\

		18 & medium number density & $q_1=2.5$, $q_2=3.5$ & 0.0012 & 0.0019 & 0.0031 & 0.0051 & 0.0075 & 0.011\\

		19 & shallow phase function & $G=1$ & \nodata & \nodata & \nodata & 0.82 & \nodata & \nodata\\

	\end{tabular}
    \caption[]{The number of ICs the LSST could observe in its lifetime for various different cases as a function of limiting magnitude. $N_{LSST}$ is the number of ICs the LSST could observe in 10 years at the listed limiting magnitude. Each row changes a few different parameters for different situations showing how $N_{LSST}$ changes. Due to the Monte Carlo nature of our analysis, each value has approximately a $\sim$10\% statistical uncertainty. The parameters and their nominal values are defined in Table ~\ref{tbl:params}. Note that the uncertainty in these parameters imply that the number of ICs detectable by LSST ranges by orders of magnitude. \label{tbl:results}}
	
	\end{center}
\end{table*}

In order to consider the effects of individual parameters, we begin with a ``nominal'' model using the parameters in \ref{tbl:params} and then consider the change in $N_{LSST}$ due to changes in particular parameters. The results for these models are shown in Table \ref{tbl:results}. We then describe a simple set of linear equations that can be used to extrapolate our results to a much wider variety of different input parameters than shown here. Finally, we consider the astrometric signature of detectable ICs. Though the unknown parameters of the IC population affect $N_{LSST}$ by orders of magnitude, we show specific values here to demonstrate the comparison between different models. 

\subsection{Nominal Model}

As discussed above, our nominal model uses standard (though uncertain and sometimes controversial) values for the input parameters, \emph{except for the size distribution, where we use the most optimistic case}. Therefore, ``nominal'' should not be misconstrued as ``best guess'' and the nominal case is designed for comparison to M09. The specific values chosen are given in Table \ref{tbl:params}. It uses the mass density of M09 and the size distribution parameters ($q_1, q_2, r_b$) that give the \emph{maximum number of small objects}. We use the \citet{Sosa11} radius-brightness relation and otherwise assume typical photometric parameters. 

Most comets seen in the inner solar system have a minimum nucleus radius of order 1 km. A common explanation for the lack of small comet nuclei is that these objects readily disintegrate and, effectively, evaporate, before they reach the inner solar system in a coherent form \citep[][]{2002Sci...296.2212L}. ICs of this size may also be destroyed by the same mechanism in previous close stellar passages. Therefore, we consider it unrealistic to consider active ICs with nuclei sizes below 0.1 km, and choose our nominal model to have a minimum IC nucleus radius of $r_{min} = 0.1$ km. Even this is very optimistic, as comet nuclei smaller than $\sim$1 km may be far rarer than the size distribution would suggest. 
Based on this model, we expect LSST to be capable of observing on the of order 1 detectable IC over its 10 year lifetime (Table \ref{tbl:results}, Line 1). There is significant uncertainty and significant optimism in this result, but it is plausible for LSST to detect an interstellar comet. These conclusions are discussed further in $\S$\ref{sec:discussion}.

\subsection{Difference Compared to M09}
Although we cannot exactly reproduce the conditions used in the analytical estimates of M09, we can approximate this model by using equivalent parameters: turning off the mass of the Sun (which removes gravitational focusing), turning off comet brightening, turning off the phase function, and only considering ICs that are at least 5 AU away from Earth. We still enforce the airmass and solar elevation constraints (a reduction of a factor of $\sim$3). This yields a detection frequency of $N_{LSST} \simeq 0.002$ (Table \ref{tbl:results}, Line 6), similar to the estimate given in M09 ($10^{-2} - 10^{-4}$). 

Considering each of these factors individually (Lines 2-5), we find that the largest effect in the increase of $N_{LSST}$ seen in our analysis is the inclusion of ICs that come closer to the Earth than 5 AU. Not only does this makes ICs brighter because they are closer, but it makes objects that are smaller detectable. Since the size distribution is steep, decreasing the minimum size of detectable ICs creates almost a 10-fold increase in the number of expected detections. This can also be seen in the modest increase gained by decreasing the nuclear bulk density; since the mass density per unit volume is fixed, this has the effect of increasing the number of objects at a certain size (Line 4).

\subsection{Varying Parameters}

A large variety of tests show that the primary determinant of $N_{LSST}$ is the number of objects at the smallest radii, unsurprising for a magnitude-limited survey of objects with steep size distributions. It also means that our results are strongly sensitive to parameters which affect the number density at the smallest radii. For a fixed number density, changing the size distribution slopes $q_1$ and $q_2$ strongly affects the number of objects at the smallest sizes (Figure \ref{fig:m1correction}) and $N_{LSST}$ correlates with these changes. Similarly, increasing the total mass density of ICs to, for example, the overly optimistic estimate of \citet{McGlynn89}, increases the detection frequency significantly to $N_{LSST} \approx 6000$. 

We illustrate this point by changing the minimum detectable radius in our ``nominal model'' from 0.1 km to 0.2, 0.5, and 1 km in Lines 8-10, which decreases $N_{LSST}$ from 0.57 to 0.39, 0.22, and 0.14, respectively. The use of different minimum radii mimics that of brighter limiting magnitudes. 

A much more realistic case than the ``nominal'' values above draws from our understanding of solar system comets. This uses a differential nuclear radius distribution of $q_1 = 2.92$ from comets \citep{2011MNRAS.414..458S} coupled with the fact that comets with sizes smaller than $\sim$1 km are unusually rare (e.g., $r_{min} = 1$ km). Although they are not as important, we also use $q_2 = 4.5$ and $n_{pre} = n_{post} = 4$ (to mitigate issues with the $H_c$-$r$-$n$ distribution mentioned above). With these more realistic parameters, $N_{LSST}$ becomes 10$^{-3}$, much lower than the nominal case because of the less favorable (but more realistic) size distribution and cut-off. Using $r_{min} = 0.1$ would still only give $N_{LSST} \simeq 0.005$. 

\begin{figure}
 \centering
 \includegraphics{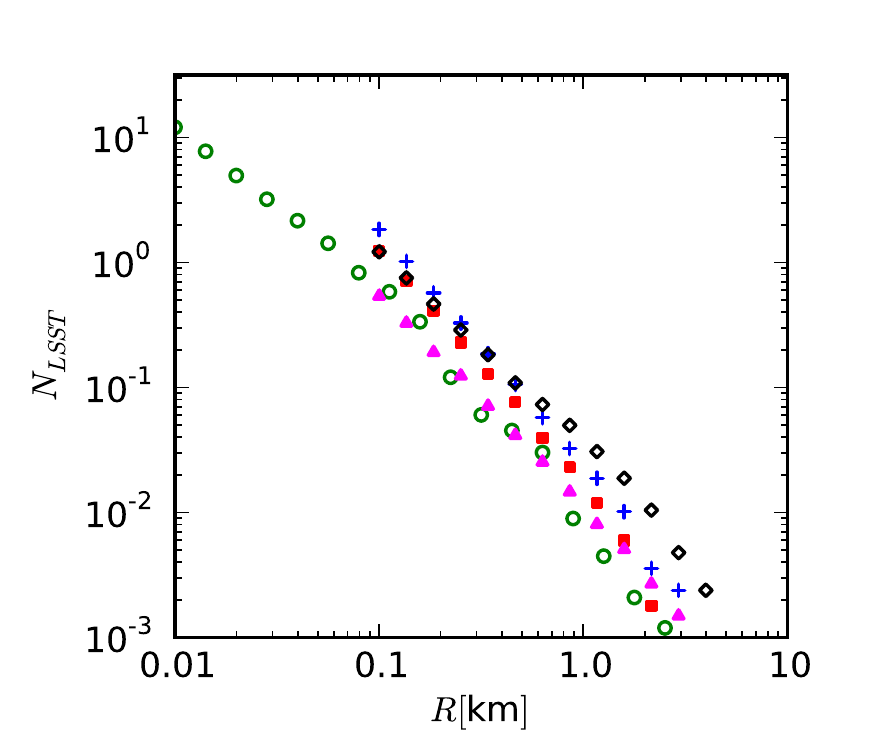}
 \caption[Comet brightening has a significant impact on the number of visible ICs.]{Comet brightening has a significant impact on the number of visible ICs. The y-axis is showing the differential number of detections over the 10-year LSST baseline (compared to the values in Table \ref{tbl:results} which are the cumulative number of detections). Here we can see the effect of the different parameters $b_1$ and $b_2$ for comet nuclei size distribution cases compared to the asteroid case. The nominal parameters are used, including extremely optimistic radius distribution differential power law slopes ($q_1 = 3.5$). The markers correspond to the different comet brightening cases and the asteroid case as follows: asteroid (green, $\circ$), Kresak 1978 (blue, $+$), Bailey \& Stagg 1988 (red, $\Box$), Weissman 1996 (magenta, $\triangle$), and Sosa \& Fernandez 2011 (black, $\diamond$). Statistical uncertainties due to Monte Carlo approximation errors are of order $\sim$10\%, which accounts for some of the variability shown. These are the numbers of detectable ICs before the observability criteria (airmass $<$ 2, solar elevation $<$ -18$^{\circ}$) are applied; requiring the objects to be observable decreases the frequency significantly, especially in the asteroid case.\label{fig:m3_cb}}
\end{figure}

Given the large uncertainties in the most important parameters, we have determined which parameters can be considered as minor effects. Based on our simulations, gravitational focusing, the specific photometric phase function, the chosen bulk nuclei density, the intrinsic IC velocity dispersion, the direction and velocity of the Sun's interstellar motion, and the different radius-brightness relations affect $N_{LSST}$ at the factor of $\lesssim$2 level only (see Table \ref{tbl:results}). Figure \ref{fig:m3_cb} shows the effect of the different size-brightness relations for comets and asteroids (see Table \ref{tbl:cb}), before observability criteria are enforced.

\subsection{Asteroid Case}

Depending on the parameters of planet formation, both rocky and icy bodies can be ejected into interstellar space \citep[e.g.,][]{1997ApJ...488L.133W, 2015MNRAS.446.2059S}. If we consider the case of interstellar asteroids (or dead/inactive comets) that are not subject to disintegration, then we can probe to much smaller sizes where the objects are much more frequent ($r_{min} = 0.01$). In this case, we move from a specific comet brightening law to using a particular albedo, nominally 0.06. However, we lose the advantage given by comet brightening. Considering the interstellar asteroid case down to a size of 10 meters shows that these two effects roughly cancel with $N_{LSST} \approx 0.9$. (Table 1, Line 12). Figure \ref{fig:m3_cb} shows how the comet radius-brightness relation and the asteroid case compare, though it is important to remember that these assume an overly optimistic value for the differential size distribution power law index. Note also that we are not considering a bimodal model of ICs, but rather assuming that the entire mass density is either in the comet or asteroid cases. 

As discussed below, interstellar asteroids of these sizes are moving very rapidly on the sky. Such objects will create a trail of significant length, even in LSST's short 15 second exposures. To account for this trailing, we adjusted the brightness of these objects down according to how long their trail would be compared to the expected FWHM of LSST detections (2''), effectively approximating a surface brightness of the trail as if it were a point source. This amounted to a small reduction of $N_{LSST}$ down to 0.75, suggesting that trailing is just beginning to become important. An optimal detection algorithm would search for statistically significant trails, even if they fell below the point source detection limit \citep[e.g.,][]{2012PASP..124.1197V}. In the case where $r_{min} \approx 0.001$ km and $N_{LSST} \approx 10$, such an algorithm would be essential, though there are unavoidable trailing losses that presumably mitigate the ability of detecting $\sim$1 meter interstellar asteroids. 

\subsection{Limiting Magnitude}

Changing the limiting magnitude modifies the typical distance at which the smallest ICs can be detected; the larger limiting magnitude of a deeper survey increases the volume where small ICs can be detected. In practice, the effect on $N_{LSST}$ is not as strong as might be expected, due to the combination of actual IC motion with respect to the Earth and the Sun, comet brightening effects, and our requirement that ICs be greater than $r_{min} = 0.1$ km. Interestingly, shallower surveys have non-negligible sensitivity, suggesting that existing surveys can place interesting upper limits on the frequency of ICs, as seen in the work of \citet{Francis05}, \citet{2004DPS....36.4008M}, and \citet{2014acm..conf..149E}. However, keep in mind that $N_{LSST}$ assumes a 10 year survey duration that is 100\% efficient at finding any detectable IC brighter than the limiting magnitude, so existing ground-based surveys should be much less sensitive.

Another important consideration is that our frequency of detections assumes that the IC must be detected in a single image or exposure. It is possible that ``shift-and-stack'' techniques \citep[e.g.][]{2010PASP..122..549P, 2015AJ....150..125H} can use the same survey to effectively reach $\sim$2 magnitudes deeper, which would certainly improve the detection rate as seen in Table 1.

The change in $N_{LSST}$ as a function of magnitude suggests that it is better to spend survey time to go ``wide'' than it is to go deeper, all else being equal. This is partly due to the fact that the IC population is continuously replenishing due to solar motion through the Galaxy. However, the rarity of ICs means that even a survey wide enough to cover the whole sky every three days will still need to go deep in order to have a clear chance of detecting ICs. It also suggests that LSST's use of ``Deep Drilling Fields'' which are more heavily observed to reach a deeper co-added magnitude, are not likely to help unless combined with a shift-and-stack technique. 

The asteroid case is much more sensitive as a function of magnitude. Very deep searches that are sensitive below $r_{min} = 0.01$ km could significantly benefit from going deeper, but would have to deal with larger amounts of trailing. A new technique which combines high-speed cameras with the shift-and-stack technique has successfully discovered Near Earth Objects (which we show below would have similar rates of motion as ICs would have) as small as $\sim$8 meters \citep{2014ApJ...792...60Z,2014ApJ...782....1S}. However, as ICs are incredibly more sparse than NEOs, this method would have to be scaled up by orders of magnitude before ICs would be detected.

\subsection{Enabling Extrapolation to Other Values of Simulation Parameters}

Through efficient programming and data control techniques, we were able to complete a full simulation of billions of ICs in, typically, several minutes, allowing for the computation of many possible models (Table \ref{tbl:results}). So that future studies can utilize and extrapolate our results, we have run several models with large variations in one or more parameters in order to see how $N_{LSST}$ changes. We have found that the following multi-linear model gives an accurate estimate. However, we again caution that the true values of most relevant parameters are not well known and that the prediction of the number of ICs detectable by LSST or other surveys ranges over orders of magnitude. 

Let $\log N = \mu(\vec{\theta}) \times log(r) +  \beta(\vec{\theta})$
where $N$ is the (differential, not cumulative) number of detectable ICs \emph{per year} at V=24.5 where $r$ is the nuclear radius of the ICs in km. To go from $N$ to $N_{LSST}$ requires multiplying by 10 (due to the 10-year baseline) and removing objects that are not observable (e.g., airmass < 2, solar elevation < -18$^{\circ}$), which is a factor of $10^{0.5-2}$ depending on the specific populations. 

At the small radius end of the distribution, the relationship between $\log N$ and $\log r$ is approximately linear (e.g., Figure \ref{fig:m3_cb}). In that regime, we can describe the relationship between them using a slope and y-intercept, $\mu$, and $\beta$, which are each themselves linear functions of the parameters $\vec{\theta}$. We describe the values of $\mu$ and $\beta$ as a linear combination of these parameters, $\vec{\theta} = (q_1, q_2, b_1, b_2, v_0, \rho, n_{pre}, n_{post}, \text{and} \log_{10}(m_{total}))$ (in the units given in Table \ref{tbl:params}), plus a constant. The values of $\beta$ are somewhat representative of the importance of this parameter, e.g., $q_1$ is much more important than $v_0$. Note that $r_{min}$ is not included because the result gives $N$ as a function of size, which can then be evaluated at the desired small size cutoff. 

\begin{table}
\begin{center}
 \begin{tabular}{l|l|l}
    & $\mu$ & $\beta$ \\
	\hline
 	constant & 2.758 & -30.399\\
	$q_1$ & -0.2364 & 3.0801\\
	$q_2$ & -0.4544 & -1.167\\
	$b_1$ & -0.28 & -2.17\\
	$b_2$ & -0.365 & -0.802\\
	$v_0$ & 0.0029 & 0.0043\\
	$\rho$ & 0.38 & 0.054\\
	$n_{pre}$ & 0.0978 & 0.275\\
	$n_{post}$ & -0.48 & -1.166\\
	$\log_{10}(m_{total})$ & 0.0022 & 0.987\\
 \end{tabular}
 \caption{Coefficients needed for extrapolation to other values of the simulation parameters. The frequency of IC detections per year above a brightness of V=24.5 ($N$) can be given by a power law for small values of the comet nucleus radius ($r$), as seen in Figure \ref{fig:m3_cb}. We can then define $\log N = \mu(\vec{\theta}) \times log(r) +  \beta(\vec{\theta})$ where $\mu$, and $\beta$ are each themselves linear functions of the parameters $q_1, q_2, b_1, b_2, v_0, \rho, n_{pre}, n_{post}, \text{and} \log_{10}(m_{total}))$ (in the units given in Table \ref{tbl:params}), plus a constant. The number of significant figures displayed helps to ensure an accurate estimate. The values of $\beta$, which are related to the normalization of the power law, are somewhat representative of the importance of this parameter, e.g., the differential size distribution slope for small values ($q_1$) is much more important than the velocity dispersion ($v_0$). Note that $r_{min}$ is not included because the result gives $N$ as a function of size, which can then be evaluated at the desired small size cutoff. \label{tbl:lin}}
\end{center}
\end{table}

The coefficients of each of the parameters in $\vec{\theta}$ in this linear model are given in Table \ref{tbl:lin}. They are obtained by performing a least squares fit for different runs of the simulation. A wide range of values for each parameter is used including varying multiple parameters at once, so as to make the model applicable to values beyond what is listed in Table \ref{tbl:results}. Although clearly not all of the figures listed are significant, using these values yields a correlation between this multi-linear approximation and the full model simulation with a high Pearson correlation coefficient ($R^2 > 0.95$). 

To illustrate, we can very unrealistically set each of these values equal to 10, which would give $\mu = 2.758 - 2.364 - 4.544 - 2.8 ... + 0.022 = -10.571$ and $\beta = -30.399 + 30.801 -11.67 ... + 9.87 = -39.445$ which implies that $N = 10^{-39.445} r^{-10.571}$, which can then be related to $N_{LSST}$ as discussed above. 

Using these results, future studies should be able to generate their own expectations for the number of detectable ICs by large scale surveys covering a wide range of observational and theoretical parameter space.

\subsection{Representative Orbits of Detectable ICs}

We have carefully created a model of the most detectable ICs under realistic observing conditions. Here, we present the orbital parameters of a representative sample of the most detectable ICs and interstellar asteroids. Table \ref{tbl:orbs} shows these parameters for 20 ICs and 20 interstellar asteroids drawn from the nominal case. 

It also serves as an ideal test population for future IC detection algorithms, similar in spirit to the Pan-STARRS Synthetic Solar System, which also contained an estimated interstellar object population \citep{2011PASP..123..423G}.

In Table \ref{tbl:orbs}, we show the radius for the IC from our simulation; this is a randomly-selected list of typical detectable IC properties from a Monte Carlo run that simulated a large population of ICs. As discussed above, the most detectable ICs typically have sizes just larger than the minimum detectable radius $r_{min}$ (here 0.1 km), since these are most abundant. 

Detected ICs have perihelia near the Earth's orbit, but detected interstellar asteroids tend to have perihelia less than 1 AU. The distribution of orbital angles is effectively isotropic. ICs with lower incoming velocities are slightly favored, as expected from the enhancement due to gravitational focusing. However, they always have significant excess velocities, so will be clearly distinguishable from near-parabolic comets with $v_{\infty} \lessapprox 0$. This reinforces our caution of interpreting C/2007 W1 (Boattini) as a true interstellar comet ($\S$ 2.2). 

\begin{table*}
 \centering
\begin{tabular}{llllllll}
    Radius $(km)$ & a (AU) & e & q (AU) & i ($^{\circ}$)& $\Omega$ ($^{\circ}$)& $\omega$ ($^{\circ}$)& $v_{\infty}$ (km/s)\\
    \hline
    0.11&-0.72625&2.4336&1.0412&60.728&109.24&331.18&34.86\\
    0.18&-6.1602&1.1571&0.96798&165.16&17.084&279.55&11.97\\
    0.21&-0.69227&5.5729&3.1657&121.13&307.16&167.01&35.70\\
    0.11&-4.8327&1.2481&1.1988&35.800&132.42&63.724&13.51\\
    0.12&-0.99266&2.2125&1.2036&83.129&355.76&118.45&29.81\\
    0.11&-0.43832&4.1940&1.4000&98.972&265.91&142.86&44.87\\
    0.12&-1.7647&1.7781&1.3747&55.182&8.6212&331.63&22.36\\
    0.20&-0.51516&4.8583&1.9876&49.244&293.03&125.17&41.39\\
    0.12&-0.51014&5.1957&2.1404&24.874&267.36&140.49&41.59\\
    0.11&-0.41419&3.8027&1.1608&51.671&293.28&115.89&46.16\\
    0.10&-1.4211&1.1802&0.25631&35.660&227.27&120.63&24.91\\
    0.14&-0.57002&3.4843&1.4161&50.452&326.48&103.23&39.34\\
    0.11&-0.42728&4.7306&1.5940&136.89&310.14&111.82&45.44\\
    0.39&-0.38712&9.1568&3.1577&124.25&309.47&125.98&47.74\\
    0.43&-1.1024&3.5955&2.8612&76.358&221.06&56.088&28.29\\
    0.12&-0.99311&1.5962&0.59213&106.96&217.56&106.32&29.81\\
    0.11&-1.9871&1.1967&0.39085&20.689&204.23&15.098&21.07\\
    0.15&-0.46875&3.5816&1.2101&67.981&80.129&278.54&43.39\\
    0.11&-1.5640&1.0915&0.14310&105.47&78.959&217.24&23.75\\
    0.11&-0.39364&6.9736&2.3515&22.201&207.98&163.29&47.34\\
    \hline
    0.032&-0.68023&1.9241&0.62862&124.65&156.97&258.53&36.02\\
    0.069&-1.6503&1.6707&1.1068&75.355&7.4526&75.896&23.12\\
    0.015&-3.5974&1.0529&0.19034&68.521&54.221&124.26&15.66\\
    0.017&-0.95701&1.0238&0.022733&6.1824&359.55&229.32&30.36\\
    0.041&-0.53558&2.2423&0.66536&167.17&258.52&77.446&40.59\\
    0.053&-0.70879&2.1711&0.83010&108.78&315.23&149.74&35.28\\
    0.11&-4.1081&1.4083&1.6775&19.809&170.96&268.51&14.66\\
    0.034&-0.42086&2.6194&0.68155&114.89&304.18&117.75&45.79\\
    0.012&-0.90621&1.5180&0.46943&86.207&249.46&94.273&31.20\\
    0.033&-1.0996&1.4696&0.51641&109.64&209.49&81.180&28.33\\
    0.018&-0.45199&2.3399&0.60561&94.492&296.68&110.37&44.18\\
    0.050&-1.0242&1.0939&0.096220&166.45&292.78&97.257&29.35\\
    0.016&-1.5610&1.4506&0.70345&169.23&16.931&136.11&23.77\\
    0.011&-0.74861&1.3012&0.22552&121.60&118.92&250.08&34.33\\
    0.028&-2.7367&1.0649&0.17765&90.813&27.294&133.56&17.96\\
    0.016&-5.6296&1.1988&1.1192&120.03&342.69&2.0826&12.52\\
    0.017&-6.2956&1.0560&0.35259&137.30&37.675&107.71&11.84\\
    0.11&-0.50025&3.9496&1.4755&167.67&243.17&63.798&41.91\\
    0.018&-4.3754&1.0949&0.41516&83.139&291.61&273.74&14.20\\
    0.29&-1.0272&2.6104&1.6542&39.742&280.81&157.18&29.31\\

 \end{tabular}
 \caption{\footnotesize Orbital properties of typical detectable interstellar comets (first 20 entries) and interstellar asteroids (last 20 entries).\label{tbl:orbs} Standard hyperbolic orbital elements in the ecliptic heliocentric reference frame are used: $a$ is semi-major axis, $e$ is eccentricity, $i$ is inclination relative to the ecliptic, $\Omega$ is the longitude of the ascending node, $\omega$ is the argument of periapse, and $v_{\infty}$ is the excess velocity at infinity.\normalsize}
\end{table*}

\subsection{Astrometric Analysis}

As input to future observational efforts, we have taken the set of the most detectable ICs shown in Table \ref{tbl:orbs} and studied their astrometry for the purpose of orbit inversion. We estimate the amount of observational data required
to determine that an object is an IC and to secure its orbit so
that follow-up observations can be obtained. 

\begin{figure}
 \centering
 \includegraphics{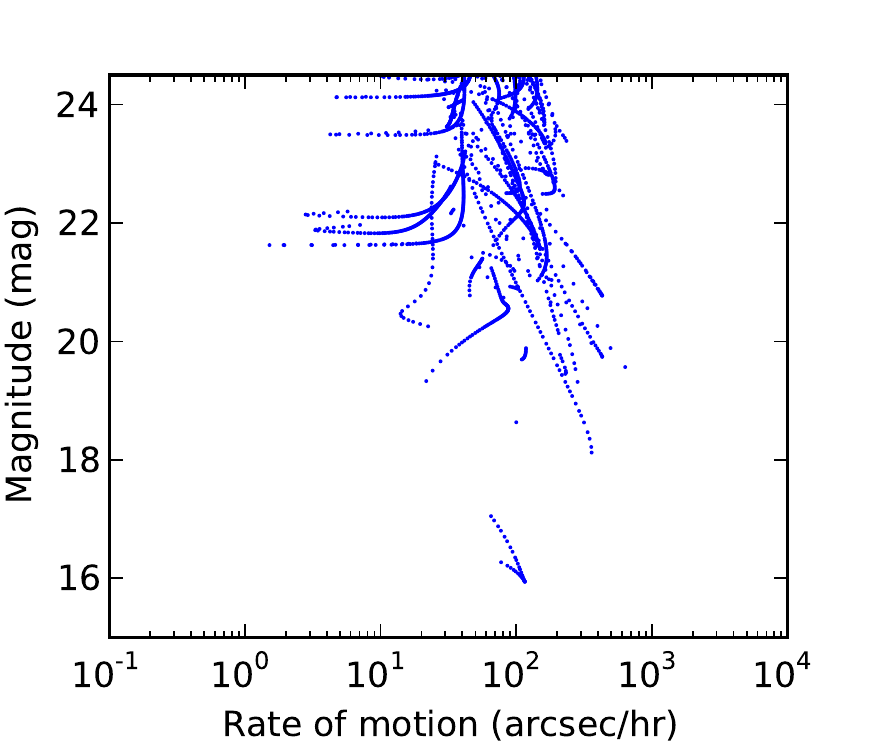}
 \caption[]{Distribution of sky rate of motion vs. magnitude for detectable ICs. In this case, the ICs were given zero velocity dispersion. The nominal and other models are similar. For individual objects (except for ICs which are observed near their apparent fixed points), an increased brightness and rate are correlated due to a closer approach to Earth.\label{fig:mag_rate}}
\end{figure}

To set the stage, Figure \ref{fig:mag_rate} shows the rate of motion of ICs the sky as a function of apparent brightness. ICs have rapid apparent motion, with typical rates of hundreds of arcseconds per hour, comparable to near-Earth objects (NEOs).

First, we generate
LSST-like synthetic astrometry for these ICs. Then the resulting astrometry is fed into a statistical orbit
computation algorithm which provides an orbital solution. Finally we
compute ephemerides based on the orbital solution. The software tools
(less the shell scripts) are available in the OpenOrb package
\citep{2009M&PS...44.1853G}
\footnote{https://github.com/oorb/oorb}. This code is different from the one developed to identify detectable ICs. Due to the difference in Earth's orbital position and a few other small details, not all of the ICs in Table \ref{tbl:orbs} are detected (at our magnitude cutoff of V=24.5) in these astrometric simulations, but this does not affect the results. This completely separate code also finds most of the ICs in Table \ref{tbl:orbs} as detectable, partially validating our simulations. 

For determining the astrometric orbits, we
use the two-body approximation, no non-gravitational effects, JPL's DE405 planetary ephemerides, and
a geocentric observer (as opposed to the topocentric observer used above). These approximations do not affect the astrometric results. 

Synthetic astrometry is generated by propagating the ICs
through their perihelion passage and recording their (RA,Dec)
coordinates twice with an interval of about 15 minutes every three
days if the apparent $V$-magnitude $V<24.5$, the solar elongation
$\epsilon_\odot>45\deg$, and the lunar elongation
$\epsilon_\mathrm{Moon}>45\deg$. Random Gaussian noise with
$\sigma=0.1''$ is added to the coordinates to mimic astrometric
uncertainty, though this may be an overestimate for a fully-calibrated LSST.

We compute an orbital solution using the statistical ranging method
\citep{2001Icar..154..412V,2005Icar..179..109G} for each night that
an object is detected twice. The process thus resembles the
operation of a real survey where the orbital uncertainty of a given
object gradually diminishes as more astrometry is obtained. The
ranging method provides the full non-linear orbital-element
probability-density function (PDF) based on the synthetic
astrometry---in practice a cloud of weighted orbital solutions that
reproduce the synthetic astrometry within the limits set by the
astrometric uncertainty. Using the PDF we assess whether elliptical
solutions can be ruled out based on the synthetic astrometry.

The most obvious characteristic that separates an IC from its
solar-system counterparts is its hyperbolic orbit with respect to the
Sun. The typically high inclinations of ICs (Table \ref{tbl:orbs}) could also be utilized but
this would possibly lead to a confusion as high inclinations and even
retrograde orbits are known to exist for both near-Earth comets and
near-Earth asteroids originating in the solar system
\citep{2012ApJ...749L..39G}. To unambiguously
determine whether an object is an IC we use the criterion that the
minimum eccentricity within the orbital-element credibility region has
to be greater than unity. In what follows, the credibility region
encompasses 99.73\% of an orbital solution's total probability mass
thus corresponding to the $3\sigma$ limit of a one-dimensional
Gaussian distribution. In astrometric analyses of solar system bodies, we have found that
single night arcs can always be fit with a hyperbolic orbit model, therefore, we require
at least 2 pairs of detections spaced by 3 nights (in our mock LSST-like cadence) before 
considering a hyperbolic detection secure. It is interesting to note that while all solar system 
objects appear to be moving retrograde at opposition due to the Earth's faster orbital velocity, 
ICs are an exception to this rule. Even at opposition, they can have a prograde motion, as a result of their excess velocity. 

\subsection{Comet Model}

Each IC is typically detected
tens or even hundreds of times during its perihelion passage. They are typically first ``discovered'' in the inner solar system at a heliocentric distance of $\sim$6 AU. The minimum and median distance to a detectable IC are $\sim$1.5 and $\sim$3 AU, respectively. If there were no lower limit on the size of an IC, i.e., if we allowed $r_{min} \lesssim 0.1$ km, then we would detect a larger population of smaller bodies closer to the Earth (see the astrometric results for the asteroid case below).

Most ICs require 2--3 independent nights of observations during a timespan of 4--7 days to ensure that a hyperbolic orbit is the only
viable solution. This rapid identification is due to the high rates of motion seen for typical ICs ($\sim$200 arcseconds/hr or $\sim$1 deg/day), allowing for precise astrometric constraints over a short observational arc. The rapid motion is partially due to the higher orbital velocity experienced by unbound objects, but is also strongly due to the fact that a magnitude-limited survey will identify nearby objects that will have rapid apparent motion due to parallax from the Earth's motion. When the data are insufficient for a strong classification, ICs are usually confused with Aethra asteroids (Mars crossers). 

In order to facilitate follow-up observations we
compute the maximum sky-plane uncertainty 3 days and 14 days after the
last detection. We assume that, e.g., photometric, spectrometric and
polarimetric follow-up observations are feasible when the ephemeris
uncertainty is below $30$" 3 days after the last observation. When
this limit has been reached it essentially guarantees that the
uncertainty does not grow with time as the ongoing survey
provides additional astrometry every 3 days. Similarly, we assume that
Target-of-Opportunity-type observations can be planned when the
ephemeris uncertainty is less than one degree within 2 weeks after the last
detection. Most objects require 3 nights of observations spanning 7 days for
the above criteria on the ephemeris uncertainty to be fulfilled. That is, the timeframe for establishing that an object is an IC is the same as the time frame for establishing its orbit sufficient for follow-up. 

The
ephemeris uncertainty drops quickly with increasing astrometry: at the
time of fulfilling the above criteria the typical 3-day ephemeris
uncertainty is only a few arcseconds, a few tens
of arcseconds for the 14-day ephemeris uncertainty, and less than
0.2${\circ}$ 30 days into the future. 

\subsubsection{Asteroid Model}

The astrometric results for ICs are probably applicable for a wide range of possible populations. However, interstellar asteroids are only detected in abundance when much smaller objects are seen and these $\sim$10-meter objects must necessarily pass very close to the Earth to be detected by LSST. As this can strongly affect the astrometric solutions, we repeated the above analysis for the asteroid population given in Table \ref{tbl:orbs}.

Unsurprisingly, the typical geocentric discovery distance was nearly always less than 1 AU and the number of detections per perihelion passage was much lower, usually less than 10. These smaller bodies require more favorable observational circumstances to be brighter than the limiting magnitude. 

Otherwise, the asteroid case was similar to the IC case: 2--3 nights spanning 4--7 days was sufficient to identify objects as hyperbolic and to have an ephemeris uncertainty small enough for recovery and Target-of-Opportunity observations. When the tracklet was too short to securely identify the object as interstellar, the most common classification was as an Apollo NEO. 

\subsection{Possible Discrete Source of ICs} \label{discrete}

Throughout the analysis thus far, we have been considering an IC population that is completely isotropic. However, it is possible that the population of ICs is more heterogeneous. If we were passing through the Oort cloud of another star, for example, the IC density would go up significantly \citep{Stern1987185}. Recent stellar passages may be effective at temporarily stripping the Oort clouds of other stars, eventually resulting in an anisotropic IC source \citep[e.g.,][]{1999MNRAS.304..579Z}.

ICs from individual systems are ejected primarily at inclinations less than $\sim$30$^{\circ}$ with respect to the invariable plane \citep[e.g.,][]{1989Icar...82..402D}. However, these planes are oriented isotropically, so that the IC ``luminosity'' of a planet forming system is similar (proportional to the inverse square of distance) to conventional photon luminosity. Throughout, we've assumed that we are searching for ICs from, effectively, the ``diffuse IC background''. Here we consider the possibility of a single dominant discrete source of ICs. 

As pointed out in $\S$\ref{sourceregion}, interstellar micrometeorites appear to have a discrete source possibly associated with edge-on debris-disk and planet hosting star $\beta$ Pictoris \citep{2000JGR...10510353B,2004ApJ...600..804M}. A similar source for ICs is plausible and may significantly enhance their frequency over the estimates of M09. 

As one of the nearest forming stars (19.4 pc), $\beta$ Pic has been extensively studied. Since the discovery of micrometeories, observers have detected a collisionally active multi-component debris disk \citep{2006AJ....131.3109G,2012Natur.490...74D} and a directly imaged planet \citep{2010Sci...329...57L,2012A&A...542A..41C} that was earlier predicted by theorists \citep[e.g.,][]{2007A&A...466..389F}. There is every reason to expect that $\beta$ Pic is actively ejecting ICs, some of which are currently passing through our solar system; the larger brethren of the already detected interstellar micrometeorites. (It's systemic radial velocity is $\sim$20 km s$^{-1}$ and therefore cannot be a source of C/2007 W1, which was verified by direct backwards integration (P. A. Dybczyński, pers. comm.).)

To investigate this possibility, we considered a model where the ICs had no intrinsic velocity dispersion and thus come streaming in due only to the Sun's relative motion. This simulates what a single population of ICs from a discrete source may look like. Gravitational focusing can cause enhancements of interstellar particles at the antapex of the Sun's motion, and we did detect a weak spatial clustering of IC detections in this case. (Our nominal model showed no spatial clustering, as expected when the velocity dispersion of the ICs is larger or comparable to the solar velocity.) 

The lower relative velocity also increases the importance of gravitational focusing and results in a slightly larger number of detectable objects to $N_{LSST}$ of 0.61, all else being equal. A ``high IC luminosity'' discrete source may also enhance the mass density of ICs, which is not included here, but could easily be a significant effect. 

It is tantalizing to note that if an IC is detected and its orbit recovered, backwards integration over several million years could reveal a very specific location for its original source, potentially identifying it as a planetesimal from a specific system like $\beta$ Pic \citep{2015MNRAS.448..588D}. Note that $\beta$ Pic is here used as an example; the actual IC population may come from other discrete sources, including, potentially, multiple sources. 

\section{Discussion and Conclusions}
\label{sec:discussion}

The likelihood of detecting interstellar planetesimals has had a long and varied history. A modern understanding of the properties of the IC population was recently proposed by M09. In this work, we've studied the realistic observational aspects of detecting and characterizing this unknown population. Using a numerical model that tracks the position and brightness of ICs, we estimate that LSST could detect on the order of 1 IC during its 10 year lifetime, with orders of magnitude uncertainty mostly based on the actual frequency of small ICs. The expected size distribution of objects reduces this rate to $\sim$0.001, but including the contribution from interstellar asteroids or comet outbursts or discrete sources may boost the detection by 1-2 orders of magnitude. Frankly, some optimism is required to conclude that LSST will detect even 1 interstellar object. 

While it is possible to improve our model, our results are sufficiently informative to begin the discussion of whether and how the astronomical community should conduct the search for ICs. Facing the stark realization that ICs are exquisitely rare, we can expect to find them at the threshold of detection. In a single night, they are generally indistinguishable from NEOs, asteroids, and long-period comets. Only at the time of the next LSST observations (nominally 3 and 6 days later), will it become clear that the orbit can only be fit when the eccentricity is greater than 1. They are moving rapidly ($\sim$200 arcsec/hr, $\sim$1 deg/day) and will be difficult to link between single night detections. It is also likely that there will be occasional false positive linkages between unrelated solar system bodies that initially appear to be ICs. Algorithms attempting to detect solar system bodies may choose to discard detections and/or linkages that indicate an unbound orbit. This may be an easy way to help make the solar system moving object search more tractable, even though it would throw away any ICs that could nominally be detected. If possible, we recommend that systems searching for NEOs, asteroids, and/or comets \citep[such as the Pan-STARRS Moving Object Processing System;][]{den2012a} refrain from explicitly or implicitly biasing their systems against the algorithmic detectability of ICs, though these are vastly less frequent than solar system small bodies. The Pan-STARRS MOPS, in particular, is not explicitly biased against hyperbolic orbits.

ICs convey rare and unique planet formation information; rare because ICs are so hard to observe and unique because their observations complement other methods used to study planet formation. The work needed to discover ICs is accompanied by a strong desire to follow them up with other observations, both for orbit recovery and for detailed characterization (e.g., with JWST). For this reason and based on our simulations, the ideal case is to discover and track ICs within 1-4 weeks, similar to NEOs. However, one of the strongest pieces of information gained from discovering an IC is their frequency and this could be determined in a specialized \emph{post facto} search, well after any detected ICs are recoverable or observable. Given the significant probability that LSST will not detect any ICs, such a project should be prepared to place an upper limit on the IC frequency based on a null detection. 

While it seems difficult to imagine now, we look forward to the day --- perhaps in the distant future --- that ICs are detected in such abundance that, like KBOs and exoplanets, the number of objects rapidly grows from zero to one to ten to a population so large it is hard to keep track of individual objects. It is exciting to consider what this future regime of IC studies could reveal about the formation and evolution of planetary systems in the Galaxy.

\acknowledgements
We thank Avi Loeb, Amaya Moro-Mart{\'{\i}}n, Luke Dones, Mike Jura, Priscilla Frisch, and Dan Green for discussions and suggestions that improved the manuscript. We thank Paul Weissman for very helpful reviews. NVC acknowledges support from the BYU Department of Physics and Astronomy mentoring support fund. We thank the Division for Planetary Sciences Hartmann Travel Grant Program for sponsoring a presentation of this work at a DPS Meeting. DR thanks Ed Sittler and the NASA Academy for support on an earlier related project. DR also acknowledges the support of a Harvard Institute for Theory and Computation Fellowship. MG was funded by grant \#137853 from the Academy of Finland. 

\appendix

\section{Other Methods for Detecting Interstellar Comets}

There may be other methods for detecting the signatures of interstellar comets besides standard direct photometric observation that we have assumed above. Here we briefly consider a few other possibilities, using the nominal mass density from M09. In considering the likelihood of detecting interstellar planetesimals, it is good to remember that they are expected to be many orders of magnitude less common than the small body populations of the solar system. 

While we have a basic model for comet brightening based on empirical observations, it is known that some comets occasionally undergo huge outbursts, increasing in brightness by several magnitudes (e.g., Comet Holmes). Unless ICs have some preference for these rare outbursts, the frequency of rare outbursts on rare ICs must be negligibly small. Another method for observing comets is to observe them approaching the Sun. This is very fruitful for sungrazing comets, like those from the Kreutz group, but is not likely to be profitable for searching for rare ICs since the volume of space that is very close to the Sun is too small. 

Interstellar comets would leave meteor shower trails like regular comets. Indeed, interstellar comet candidate C/2007 W1 caused a readily detectable meteor shower from its single passage through the inner solar system \citep{2011MNRAS.414..668W}. Passing ICs could go undetected but cause streams that may be intercepted by the Earth, but generally these would be a tiny fraction compared to streams caused by solar system bodies. In fact, there can be hyperbolic components to meteors caused by gravitational interactions in the solar system, adding to the confusion \citep{2014Icar..242..112W}. Previous interactions between ICs and other stars would have left trails that lace and thread the galaxy, but these presumably have a short lifetime, rendering detection and characterization unlikely. 
There may be signs of IC accretion onto solar system bodies. ICs can have unusually high orbital velocities and would often create hypervelocity craters on practically any solar system surface. However, the velocities are not expected to be so high as to be otherwise inexplicable for bodies in the inner solar system. In the Kuiper belt, where typical impact velocities between Kuiper Belt Objects are only $\sim$1 km/s, an impact by an IC with a velocity of $\sim$25 km/s could produce a somewhat unusual crater, although to first order only the impact energy can be deduced from a crater and not the initial velocity. Collisions between Kuiper Belt Objects and Oort Cloud comets can have higher collision velocities $\sim$5 km/s. However, even the most favorable IC population from M09 would suggest that the largest IC to hit Pluto over the age of the solar system was $\sim$1 meter in radius, which would make a crater far too small to be detected by New Horizons \citep{2008SSRv..140...75W}. The IC accretion rate and the integrated accretion are both so much smaller than accretion from solar system material that even extreme chemical or isotopic differences would be washed away. For example, based on the highest M09 abundance, the largest IC that has ever hit the Earth (which has intercepted a volume of $\sim$60 AU$^3$ over its 4.5 GYr history) is $\sim$30 m in radius, not even Tunguska-size. The dependence on the tiny mass density of ICs and small cross-sections of the planets echo studies that conclude interstellar panspermia is very difficult, at best \citep[e.g.,][]{2003AsBio...3..207M,2004MNRAS.348...52W,2012AsBio..12..754B}. 

Some ICs are gravitationally captured by the solar system \citep{1982ApJ...255..307V}. \citet{1986AJ.....92..171T} study the ability of Jupiter to capture ICs and conclude that these captures would only occur every $\sim$ 60 MYr using the unrealistically high density of \citet{McGlynn89}. Using estimates of M09, the total volume of planetesimals captured in this way over the age of the solar system is $\sim$1 km$^3$. However, the analysis of \citet{1986AJ.....92..171T} assumed a single 20 km/s characteristic velocity for ICs and did not consider either a velocity dispersion, chaotic temporary captures, tidal disruption, or a possible discrete source of ICs. Nor did \citet{1986AJ.....92..171T} consider the post-capture orbital evolution of ICs. Since the new heliocentric orbit intersects the orbit of Jupiter, captured ICs will not generally be dynamically long-lived (and could be re-ejected). Recent captures that have not yet been destroyed or re-ejected may have unique heliocentric orbits, though the largest bodies on these orbits are probably only tens of meters in diameter or smaller, making them too small to detect except in extraordinary circumstances. Intriguingly, there are known comets (e.g., 96P/Machholz) with unusual chemical abundances and orbital parameters that have been hypothesized to have an interstellar origin \citep{2007ApJ...664L.119L,2008AJ....136.2204S}, though the expected abundance from M09 makes this very unlikely. Our model could fruitfully be expanded to include Jupiter and its gravitational influence to investigate this possibility in more detail, which we leave to future work. 

Things can be different in the early solar system. First, the initial proto-solar nebula was presumably seeded with those ICs from previous generations of planet formation that had either a low velocity relative to the nebula by chance to be captured by aerodynamic drag. These are the larger components of ``pre-solar grains'' found in meteorites. Comparing the M09 estimated density of ICs with the density of the ISM, suggests that roughly $10^{-6}$ of the material (by mass) in the proto-solar nebula may come from ICs. This may be sufficient to affect some processes of planet formation, which usually assume a pure dust and gas initial composition; an extreme example is that ICs may serve as seed particles to break the grain-grain bouncing growth barrier \citep{2012A&A...540A..73W}. In addition, a miniscule fraction of primitive small bodies in the solar system (and the $\sim$50000 meteorites in various collections) may actually have an extra-solar provenance, though it would be very difficult to prove this conclusively. Birth in a stellar cluster may significantly enhance the probability of capturing ICs into the early solar system \citep[e.g.,][]{2012ApJ...750...83P,2010Sci...329..187L}, though this would decrease any distinguishing chemical features assuming the proto-cluster to be relatively homogeneous. Indeed, it is not even clear that these would be considered ``extra-solar'' at all. 

Another possible detection method is serendipitous occultations. At first, this seems hopeless; the Kuiper belt is far more dense than the population of ICs, but there have only been clear detections of 2 small KBOs \citep{2012arXiv1210.8155S}, and even these are now severely called into question by New Horizons crater population statistics. On the other hand, presumably ICs are distributed throughout the galaxy, allowing for the volume along a line of sight to a distant object to potentially compensate for the low density of ICs. More distant objects have a larger IC ``optical depth'', resulting in something like Olbers paradox \citep[in the geometric optics limit, which does not generally apply, see][]{2010MNRAS.402L..39H}. It can be shown that the geometric enhancement in size for closer objects is exactly canceled by the smaller number of these objects in the cone-shaped line-of-sight to a distant object and that (for $q_1 > 2$), smaller ICs dominate over large ICs in optical depth. For objects even $\sim$10 kpc away using the density of M09, the covering fraction by cm-size ICs is only roughly $10^{-7}$. Extrapolating to smaller ICs becomes effectively the same as standard dust extinction towards distant objects. In any case, occultations are not a viable method for detecting the frequency of ICs or individual objects. 

It is fair to conclude that the most likely method for detecting ICs is direct optical observation of the continuously inflowing hyperbolic IC population as discussed in the main text above.

\bibliography{main}
\bibliographystyle{apj}

\end{document}